\documentclass[aps,prd,twocolumn,amssymb,eqsecnum,showpacs,showkeyes,secnumarabic,graphics,floatfix,nofootinbib,tightenlines,longbibliography,superscriptaddress]{revtex4-1}
 
\usepackage{graphicx}
\usepackage{epsfig} 
\usepackage{bm}
\usepackage{cancel}  
\usepackage{dcolumn}  
\usepackage{amsmath} 
\usepackage{hhline} 
\usepackage[dvipsnames]{xcolor}
\usepackage[utf8]{inputenc}
\usepackage{cancel}
\usepackage[dvipsnames]{xcolor}  
\usepackage{float}
\usepackage{longtable}
\usepackage[breaklinks=true,colorlinks=true,
linkcolor=blue,urlcolor=blue,citecolor=MidnightBlue,% PDF VIEW
bookmarks=true,bookmarksopenlevel=2]{hyperref}
\usepackage{mathrsfs} 
\usepackage{stackengine}
\newcommand\xrowht[2][0]{\addstackgap[.5\dimexpr#2\relax]{\vphantom{#1}}}

\begin{document}

\title{Q-balls in K-field theory}
\author{Aníbal Faúndez}\email{anibal.faund@gmail.com}
\author{Radouane Gannouji}\email{radouane.gannouji@pucv.cl}
\affiliation{Instituto de Física, Pontificia Universidad Católica de Valparaíso,
Av. Brasil 2950, Valparaíso, Chile}

\date {\today}  

\begin{abstract}
We study the existence and stability of Q-balls in noncanonical scalar field theories, $K(|\Phi|^2,X)$ where $\Phi$ is the complex scalar field and $X$ is the kinetic term. We extend the Vakhitov-Kolokolov stability criterion to K-field theories. 
We derive the condition for the perturbations to have a well-posed Cauchy problem. We find that $K_{,X}>0$ and $K_{,X}+XK_{,XX}>0$ are necessary but not sufficient conditions. The perturbations define a strongly hyperbolic system if $(K_{,X}-2\phi'^2 K_{,XX})(K_{,X}+2\omega^2\phi^2 K_{,XX}) > 0$.
For all modifications studied, we found that perturbations propagate at a speed different from light. Generically, the noncanonical scalar field can lower the charge and energy of the Q-ball and therefore improves its stability.
\end{abstract}
\maketitle 

\section{Introduction}

Q-balls are pseudolike particles that could be defined as lumps of a singularity-free scalar field with finite energy. They have been  originally discovered in \cite{Rosen:1968mfz} and independently rediscovered in \cite{Coleman:1985ki}. Contrary to solitons, they do not have a topological charge but a Noether charge based originally on the $U(1)$ global symmetry, and therefore they belong to the class of nontopological solitons. The scalar field is captured in some region of space because of nonlinear self-interaction, therefore forming a pseudolike particle carrying charge and energy. 

Q-balls can be produced via many mechanisms, which makes them very interesting in particular in cosmology. Indeed, they could be produced from inflationary models, such as natural inflation \cite{Freese:1990rb,Adams:1992bn}, where if a complex scalar field with a global symmetry is spontaneously broken, we end up with the inflaton as the goldstone boson and a naturally flat potential due to the shift symmetry. Also in supersymmetric extensions of the standard model (see e.g. \cite{Kusenko:1997zq}), Q-balls emerge naturally where the global charge could be assumed by the baryon or the lepton number. For example, the Affleck-Dine mechanism \cite{Affleck:1984fy,Dine:1995kz} uses the supersymmetric flat directions to generate baryogenesis. In this context, some of these flat directions (scalar field) can be parametrized as a complex field, which is in general a condensate of squarks, sleptons and Higgs field. This condensate can be unstable and form Q-balls \cite{Enqvist:2003gh}.

Of course, the most interesting property of Q-balls is their stability, because they could then be considered dark matter candidates \cite{Kusenko:1997si,Kusenko:2001vu}. For that reason, it will be our main focus in this paper along with some interesting properties related to their existence. The analysis of the classical stability was studied in \cite{Friedberg:1976me,Lee:1991ax} where they found that considering a Q-ball of frequency $\omega$ and charge $Q$, stability is similar to the condition $dQ/d\omega <0$. It was shown in \cite{Panin:2016ooo} that the stability of gauged Q-balls is not related to this condition. It would be interesting to see the extension of this criteria to global charge Q-balls but in modified gravity theories.

We will study three types of stability conditions that appear in the literature \cite{Tsumagari:2008bv}, namely, classical stability as we have previously mentioned, absolute stability, and stability against fission \cite{Lee:1991ax}.
%and finally using the catastrophe theory \cite{Sakai:2007ft}.

In most of the papers, a canonical scalar field is assumed, which appears naturally at low energies of various theories. But studying Q-balls in the early universe might modify this simple picture. Indeed, e.g. higher dimensions naturally produce scalar fields with nonlinear kinetic terms such as D3-brane \cite{Silverstein:2003hf} or in the context of braneworld gravity \cite{Goon:2011qf}. Also in string theory, a rolling tachyon has a Dirac-Born-Infeld (DBI) type of action \cite{Sen:2002an}. It is therefore natural to look to noncanonical scalar fields. Q-balls in the DBI type of kinetic term was studied in \cite{Kuniyasu:2016tse} along with its stability using catastrophe theory \cite{Sakai:2007ft}. In this context, we will study Q-balls in the context of a complex K-field also known as K-inflation \cite{Armendariz-Picon:1999hyi} or K-essence \cite{Armendariz-Picon:2000nqq}.

The plan of the paper is as follows. We introduce the model before discussing the stability conditions encountered in the literature. In the next section, we analyze the range of existence of the Q-balls and define the energy conditions for these solutions. Finally, we will study numerically the properties of the Q-balls before studying the equation of perturbation. We analyze the strong hyperbolicity of these equations along with the stability of the Q-ball before conclusions.

%In a second part of the paper, we will look to the gravitational influence on the existence and stability of these Q-balls. In this context, they are mostly known as boson stars. They have often been studied as black-hole mimickers, or just as compact objects and obviously as candidates for dark matter.

\section{Q-balls}

Let us consider the density Lagrangian 
\begin{align}
\mathcal{L}= K(|\Phi|^2,X)
\end{align} 
where $K$ is a generic function of a complex scalar field $\Phi$ and the kinetic term $X=-\partial_\mu\Phi\partial^\mu\Phi^*$. The equation of motion is
\begin{align}
    \nabla_\mu(K_{,X} \partial^\mu \Phi)+\Phi K_{,|\Phi|^2} =0
    \label{eq:flat}
\end{align}
where we have used the notation $K_{,A}\equiv {\partial K}/{\partial {A}}$.

%This equation can be expanded into
%\begin{align}
% K_{,X}\Box\Phi-K_{,XX}\partial^\mu\Phi\Bigl(\nabla_{\mu\nu}\Phi\partial^\nu\Phi^*+\nabla_{\mu\nu}\Phi^*\partial^\nu\Phi\Bigr)\nonumber\\
%+K_{|\Phi|^2X}\partial^\mu\Phi\Bigl(\Phi^*\partial_\mu\Phi+\Phi\partial_\mu\Phi^*\Bigr)
% +\Phi K_{,|\Phi|^2} =0
%     \label{eq:flat}
%\end{align}
%from which we observe that we have $\Phi_{,\mu\nu}$ and $\Phi^*_{,\mu\nu}$. Hence the system of equations has to be written in the following form 
%\begin{align}
%    \begin{pmatrix}
%A & -K_{,XX}\Phi^{,\mu}\Phi^{,\nu} \\
%-K_{,XX}\Phi^{*,\mu}\Phi^{*,\nu} & A
%\end{pmatrix}\begin{pmatrix}
%\partial_{\mu\nu}\Phi \nonumber\\
%\partial_{\mu\nu}\Phi^* 
%\end{pmatrix}\\
%+F(\partial_\mu\Phi,\partial_\mu\Phi^*,|\Phi|^2)=0
%\end{align}
%with $A=K_{,X}g^{\mu\nu}-K_{,XX} \Bigl(\Phi^{,\mu}\Phi^{*,\nu}+\Phi^{*,\mu}\Phi^{,\nu}\Bigr)$. The matrix defines an effective metric which determines the cone of influence for the k-field and therefore provides a condition for the hyperbolicity of the system. The determinant of that matrix should be positive....

The model admits a global U(1) symmetry with which the associated Noether current is
\begin{align}
    j^\mu= i K_{,X} \Bigl(\Phi^*\partial^\mu\Phi-\Phi\partial^\mu\Phi^*\Bigr)
\end{align}
This current is conserved $\partial_\mu j^\mu$ on-shell. The corresponding conserved scalar charge (or total particle number) is 
\begin{align}
    Q=\int {\rm d}^3 x j^0=i\int {\rm d}^3 x K_{,X}(\Phi \dot\Phi^*-\dot \Phi \Phi^*)
\end{align}
To obtain the energy, we define the canonical conjugate momenta to the variables $\Phi$ and $\Phi^*$,
\begin{align}
    &\pi_\Phi = \frac{\partial \mathcal{L}}{\partial \dot\Phi}= K_{,X}\dot\Phi^*\\
    & \pi_{\Phi^*} = \frac{\partial \mathcal{L}}{\partial \dot\Phi^*}= K_{,X}\dot\Phi
\end{align}
so the Hamiltonian density is
\begin{align}
    \mathcal{H}= \pi_\Phi \dot\Phi + \pi_{\Phi^*} \dot\Phi^*-\mathcal{L}=2|\dot\Phi|^2 K_{,X}-K
\end{align}
The energy of the system is then
\begin{align}
    E= \int {\rm d}^3 x \Bigl(2|\dot\Phi|^2 K_{,X}-K\Bigr)
    \label{eq:energy}
\end{align}
We are looking for solutions that minimize the energy for a given charge Q. For that, we define the functional
\begin{align}
    E_\omega= E + \omega \Bigr[Q-i\int {\rm d}^3 x K_{,X}(\Phi \dot\Phi^*-\dot \Phi \Phi^*)\Bigl]
\end{align}
where $\omega$ is a Lagrange multiplier which enforces the given charge Q. We have
\begin{align}
    E_\omega &= \omega Q + \int {\rm d}^3 x \Bigl[K_{X}\Bigl(2|\dot\Phi|^2-i\omega (\Phi \dot\Phi^*-\dot \Phi \Phi^*)\Bigr)-K\Bigr]\nonumber\\
    & =\omega Q + \int {\rm d}^3 x \Bigl[K_{X} |\dot\Phi-i\omega \Phi|^2+K_{X}(|\dot\Phi|^2-\omega^2|\Phi|^2)\nonumber\\
    &\qquad\qquad\qquad-K\Bigr]
\end{align}
In the case of a canonical scalar field, $K=X-V(|\Phi|^2)$, we have
\begin{align}
       E_\omega =\omega Q &+ \int {\rm d}^3 x \Bigl[|\dot\Phi-i\omega \Phi|^2-\omega^2|\Phi|^2+|\vec\nabla\Phi|^2\nonumber\\
       &+V(|\Phi|^2)\Bigr]
\end{align}
where we used that $X=-\partial_\mu\Phi\partial^\mu\Phi^*=|\dot\Phi|^2-|\vec\nabla\Phi|^2$. We can therefore conclude that for a given charge Q, the energy is minimized when $\dot\Phi-i\omega \Phi=0$ which means for $\Phi(t,\vec{x})=\phi(\vec{x})e^{i\omega t}$ \cite{Lee:1991ax}. This simple argument for the canonical scalar field cannot be easily generalized to the K-field. But we observe that in the general case, if $\Phi(t,\vec{x})=\phi(\vec{x})e^{i\omega t}$,
\begin{align}
    E_\omega =\omega Q - \int {\rm d}^3 x ~K
\end{align}
which implies that the extrema of the energy (for fixed charge) coincide with the extrema of the action. Therefore solutions of the following type $\Phi(t,\vec{x})=\phi(\vec{x})e^{i\omega t}$ extremize the energy. Even if we do not know of the existence of other solutions that could also extremize the energy functional, we will assume in the future for this paper this time-dependent phase of the solution.

For a given model, the only parameter that characterizes the energy $E$ and the charge $Q$ is the parameter $\omega$. Therefore we can consider that energy and charge are functions of $\omega$, thus differentiating the energy, and we get
\begin{align}
    \frac{{\rm d}E}{{\rm d}\omega}=\int {\rm d}^3 x\Bigl[2\omega\phi^2K_{,X}+4\omega^3\phi^4K_{,XX}\Bigr]
\end{align}
Performing the same differentiation of the charge $Q$, we found
\begin{align}
    \frac{{\rm d}E}{{\rm d}\omega}= \omega \frac{{\rm d}Q}{{\rm d}\omega}
    \label{eq:EQ2}
\end{align}
which extends to K-field results from \cite{Friedberg:1976me}. When $\frac{{\rm d}Q}{{\rm d}\omega}=0$ also $\frac{{\rm d}E}{{\rm d}\omega}=0$ which corresponds to the existence of extremum of the charge and the energy at the same time. They will correspond to the cusps in the diagram $E(Q)$. When $\frac{{\rm d}Q}{{\rm d}\omega}\neq 0$, we obtain
\begin{align}
    \frac{{\rm d}E}{{\rm d}Q}=\omega
    \label{eq:EQ}
\end{align}
which corresponds to the generic relation found for a $U(1)$ Q-ball.

\section{Stability}

Usually, three different stability criteria are discussed in the literature. The first condition considers that a given Q-ball should not decay into smaller Q-balls, sometimes referred to as stability against fission \cite{Lee:1991ax}. In that case, the stability translates into 
\begin{align}
    E(Q_1+Q_2) < E(Q_1) + E(Q_2)
\end{align}
and if taking derivatives with respect to both charges ($Q_1,Q_2$), we obtain the equivalent condition $\frac{{\rm d}^2 E}{{\rm d} Q^2} <0$; by using Eq.(\ref{eq:EQ}) it reduces to $\frac{{\rm d}Q}{{\rm d}\omega}<0$. Notice the similarity with the more generic Vakhitov-Kolokolov stability criterion \cite{VK} (or spectral stability). Of course, because of Eq.(\ref{eq:EQ2}), we could equivalently consider $\frac{{\rm d}E}{{\rm d}\omega}<0$.

The second stability criterion considers decay into free particles of mass $M=\sqrt{\frac{V''(0)}{2}}$. To avoid the decay of a Q-ball into $Q$ free particles with the rest masses $M$, we need to consider $E(Q)<M Q$. 

Finally, the last stability considers the time evolution of small perturbations, the so-called classical stability that we will analyze later.

Notice that from the catastrophe theory, a simple criteria of stability has been proved \cite{Schunck:1992}. Indeed, considering the diagram $E(Q)$, the lowest branch corresponds to the stable soliton while the upper branch is unstable. This condition will be found to be equivalent to the linear stability.

\section{Existence}
\label{section:potential}
In this section, we briefly summarize the conditions of the existence of Q-balls. These conditions are obtained by constraining the shape of the potential. 

Considering a flat spherically symmetric spacetime, and $\Phi=\phi(r) e^{i\omega t}$; Eq.(\ref{eq:flat}) becomes
\begin{align}
    &K_{,X}\Bigl(\phi''(r)+\frac{2}{r}\phi'(r)+\omega^2 \phi(r)\Bigr)+\phi'(r) X'(r) K_{,XX}\nonumber\\
    &\qquad +\phi'(r)^2 K_{,\phi X}+\frac{1}{2}K_{,\phi}=0
\end{align}
with $X=\omega^2 \phi(r)^2-\phi'(r)^2$.

Let us first consider the canonical case, namely $K=X-V(\phi)$. The  equation of motion reduces to
\begin{align}
    \phi''(r)+\frac{2}{r}\phi'(r)+\omega^2 \phi(r)-\frac{V'(\phi)}{2}=0
\end{align}
which can be written as
\begin{align}
    \phi''(r)+\frac{2}{r}\phi'(r)-V_{\text{eff}}'(\phi)=0
\end{align}
with $V_{\text{eff}}(\phi)=(V(\phi)-\omega^2\phi^2)/2$. We see that the $\omega^2$ term acts as a tachyonic contribution to the mass of the field, which will produce solitonic solutions otherwise absent for $\omega=0$. Considering only solutions with finite energy, the energy functional (\ref{eq:energy}) $E=\int{\rm d}^3x (\phi'(r)^2+\omega^2 \phi^2+V(\phi))$ implies that $(\phi,\phi') \rightarrow 0$ for $r \rightarrow \infty$ and $V(0)=0$ (we assumed $V(\phi)>0$). 

It is easier to use the analogy with a particle in Newtonian mechanics, namely replacing $\phi\rightarrow x$ and $r \rightarrow t$ which gives $ \ddot x + \frac{2}{t} \dot x +W_{\text{eff}}'(x)=0$, where $W_{\text{eff}}(x)=-V_\text{eff}(x)$. Looking for a trajectory $\phi(r)$ or equivalently $x(t)$, we need to impose $x(\infty)=0$ to obtain a finite energy solution. Therefore, the problem reduces to classifying the different trajectories of the equivalent particle giving finite energy. It is easy to show \cite{Coleman:1985ki} that we need to impose $W''_{\text{eff}}(0)<0$ and $W_{\text{eff}}(\phi)>0$ around $\phi(r=0)$. These conditions translate into
$V''(0)>2\omega^2$ as well as  $\text{min}\Bigl(\frac{V(\phi)}{\phi^2}\Bigr)\leq \omega^2$. Thus, nonrenormalizable potentials have to be considered and the simplest could be $V(\phi)=m^2 \phi^2-b\phi^4+\lambda \phi^6$. The previous constraints reduce to
\begin{align}
    0<m^2-\frac{b^2}{4 \lambda}<\omega^2 \leq m^2
\end{align}
The positivity of $m^2-b^2/4 \lambda$ is imposed by demanding that $V(0)$ is a global minimum. In this paper, we will normalize \cite{Volkov:2002aj} the parameters such as $\lambda=1$ and $b=2$ which implies $m>1$. Therefore we will consider $m^2=1.1$ which implies $0.32<\omega \leq 1.05$. The Q-ball will exist only in this range of frequencies. It is important to mention that this range will change for K-fields. For example, in a model where $K=X+\alpha X^2-V(\phi)$, we have around $r=0$, and using the condition $\phi'(r=0)=0$, $\phi''(r)+W'_{eff}(\phi)\simeq 0$ with
\begin{align}
    W'_{eff}=\omega^2\phi-\frac{m^2-2 b \phi^2+3\lambda \phi^4}{1+2\alpha \omega^2\phi^2}\phi
\end{align}
Therefore the condition $W_{eff}>0$ for some range of the scalar field, implies a different value for the minimum of $\omega$. For our parameters, we found that with good accuracy, $\omega_{min}\simeq (1+\alpha/30)/\sqrt{10}$ while $\omega_{max}$ remains unchanged.

Another important condition for the existence of the Q-ball is the nature of the differential equation. We have an equation
\begin{align}
    \Bigl(K_{,X}-2\phi'^2 K_{,XX}\Bigr)\phi''+F(\phi,\phi') = 0
\end{align}
To avoid singular points, we need to impose $K_{,X}-2\phi'^2 K_{,XX} \neq 0$. Therefore, for any model, smoothly connected to the canonical case, $K_{,X}-2\phi'^2 K_{,XX} =1$, we should impose $K_{,X}-2\phi'^2 K_{,XX}>0$. Considering the model $K=X+\alpha X^2-V(\phi^2)$, we have $1+2\alpha\omega^2\phi^2-6\alpha\phi'^2>0$. Around the origin, we have $\phi'=0$, which implies the condition $1+2\alpha\omega^2\phi_0^2>0$ and therefore large negative values of $\alpha$ will not be allowed.

\section{Energy conditions}

For these type of models, the fluid interpretation is not suitable because the kinetic term does not have a definite sign. But, it is mostly positive in the interior of the Q-ball and becomes negative near the surface of the Q-ball. Therefore, deep inside the Q-ball, we can use the hydrodynamical interpretation of the scalar field, by defining the energy-momentum tensor
\begin{align}
    T_{\mu\nu}=K g_{\mu\nu}+K_{,X} (\partial_\mu\Phi \partial_\nu\Phi^*+\partial_\mu\Phi^* \partial_\nu\Phi)
\end{align}
from which we define the energy density $\rho=2|\dot\Phi|^2 K_{,X}-K=2\omega^2\phi(r)^2 K_{,X}-K$, the radial pressure $P_r=2\phi'(r)^2K_{,X}+K$ and finally the tangential pressure $P_t=K$. These quantities can be converted into the pressure $P=(P_r+2P_t)/3$ and the shear force $S=P_r-P_t$. Notice that the energy defined from $E=\int{\rm d}^3 x T_{00}$ corresponds to Eq.(\ref{eq:energy}).

The hydrodynamical approach helps to obtain easily the energy conditions such as the strong energy condition (SEC)
\begin{align}
    K_{,X}\geq 0\,,~~
    K+(\omega^2\phi^2+\phi'^2)K_{,X}\geq 0
    \label{SEC}
\end{align}
the dominant energy condition (DEC)
\begin{align}
    K_{,X}\geq 0\,,~~(\omega^2\phi^2-\phi'^2)K_{,X}-K \geq 0
    \label{DEC}
\end{align}
the weak energy condition (WEC)
\begin{align}
    K_{,X}\geq 0\,,~~2\omega^2\phi^2 K_{,X}-K \geq 0
    \label{WEC}
\end{align}
and the null energy condition (NEC)
\begin{align}
    K_{,X}\geq 0
    \label{NEC}
\end{align}
We notice that $K_{,X}\geq 0$ is common to all energy conditions.

\section{Numerical analysis}

\begin{figure}[H]
\centering
\includegraphics[scale=0.48]{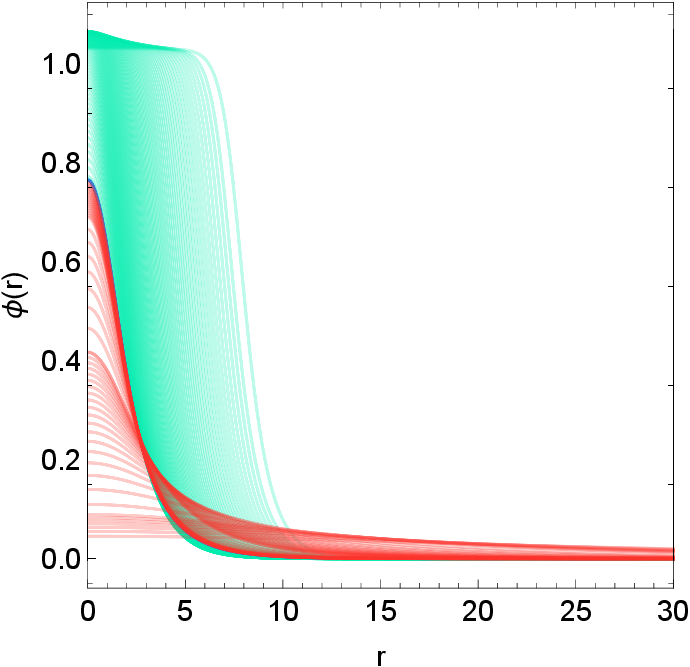}\hspace{0.5cm}
\includegraphics[scale=0.445]{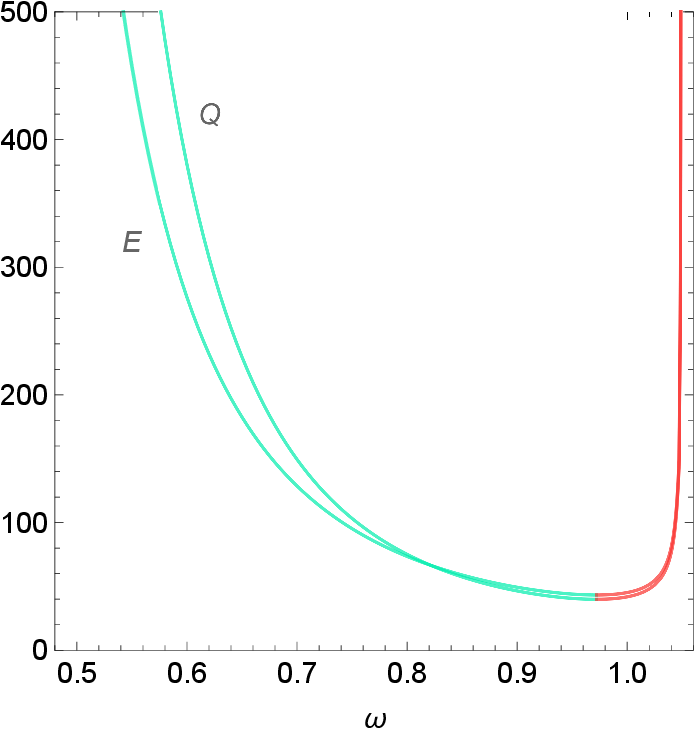}\hspace{0.5cm}
\includegraphics[scale=0.48]{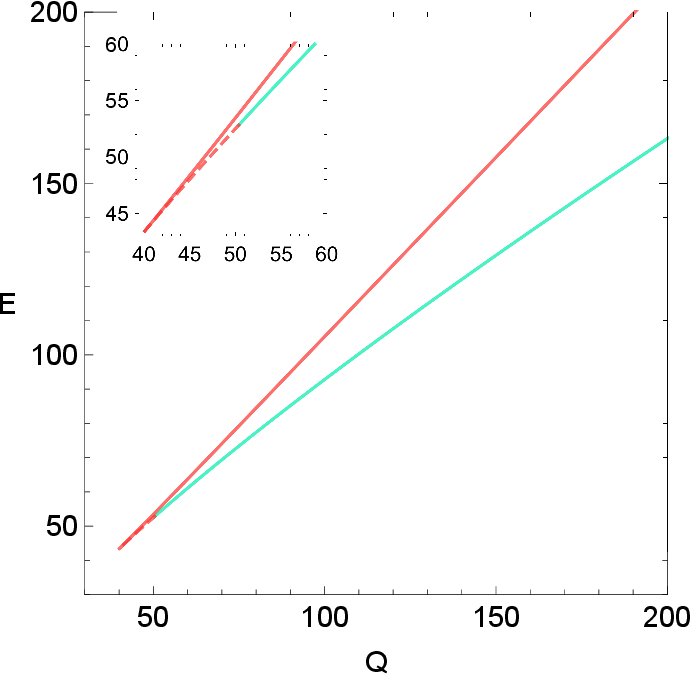}
\caption{Left: The field $\phi(r)$ is shown as a function of the radial coordinate for different values of $\omega$. For each value of $\omega$, $\phi(0)$ is adjusted such that $\phi(\infty)=0$. Center: The energy $E$ and the charge $Q$ are shown as a function of the frequency $\omega$ with the critical frequency (change of colors) defined by the condition $dQ/d\omega=0$. Right: The energy is shown as a function of the charge. For all graphics, in green we have stable configurations according to the fission stability criteria, while in red we have unstable solutions. In the first figure, the solution for the critical frequency is shown in blue and in the third graphics, we have added the decay stability criteria that is shown by a red solid line and red dashed line for the unstable solutions while the fission unstable configurations are represented only by red solid line.}
\label{Fig:standard}
\end{figure}

As we have mentioned, Q-balls are finite energy objects and therefore with a finite space extension, which imposes the asymptotic condition $\phi(\infty)=0$. Therefore we have used a shooting method for each value of the frequency $\omega$ with mixed boundary conditions $\phi'(0)=0$ and $\phi(\infty)=0$. In practice, we have integrated the system from $r=10^{-30}$ to some value, $r_{\text{max}}$, and demanded that the solution remains unchanged if we increase $r_{\text{max}}$. In Fig. \ref{Fig:standard}, we have considered the standard model $K(X)=X-V(|\Phi|^2)$ with the potential defined in Sec. \ref{section:potential}. For lower frequencies, or the thin wall limit, the scalar field is constant and at some radius (often considered as the Q-ball radius) the scalar field drops rapidly to zero, while for larger values of $\omega$, also known as the thick wall limit, the scalar field is more shallow. The latter will be unstable. In the same graphics, we have represented the energy and the charge. The energy and charge seem to diverge for the frequencies $\omega_{min}$ and $\omega_{max}$. Also $E(\omega)$ and $Q(\omega)$ reach their minimum for the same frequency, defining therefore a cusp in the energy vs charge graphics. We show also the stability conditions of the Q-balls. The stability criteria against decay is stronger than the fission stability condition. In the $(Q,E)$ plot, it is easy to determine the stable Q-ball. Indeed, for every given charge $Q$, two Q-balls exist, and the one with the smallest energy corresponds to the solution stable under fission. We will see later, that it corresponds also to the stable solution under linear perturbations. 

\onecolumngrid

\twocolumngrid

\begin{figure}
\centering
\includegraphics[scale=0.458]{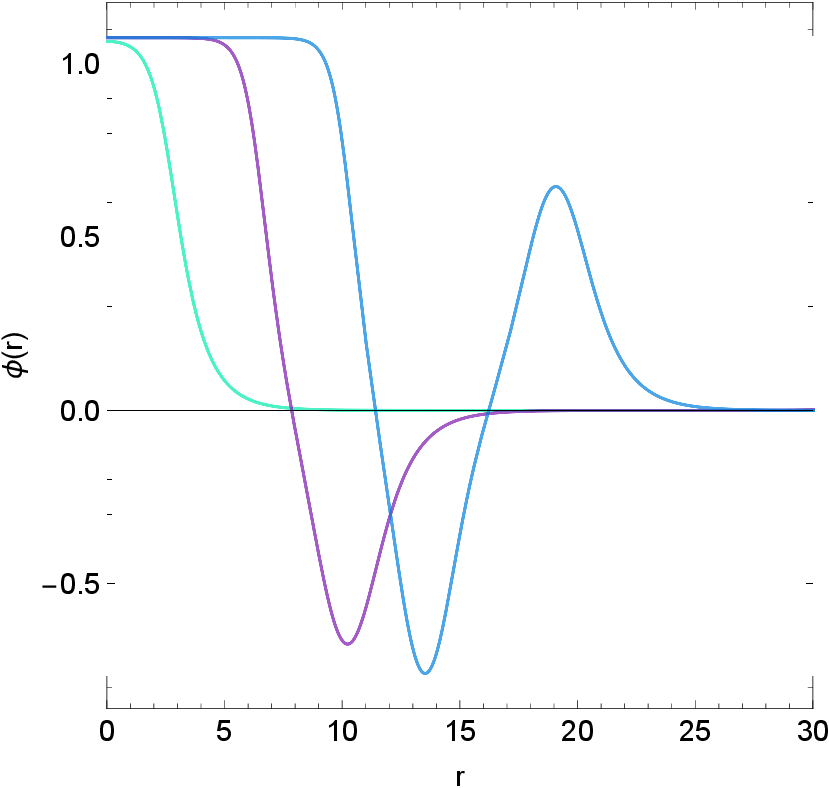}
\includegraphics[scale=0.45]{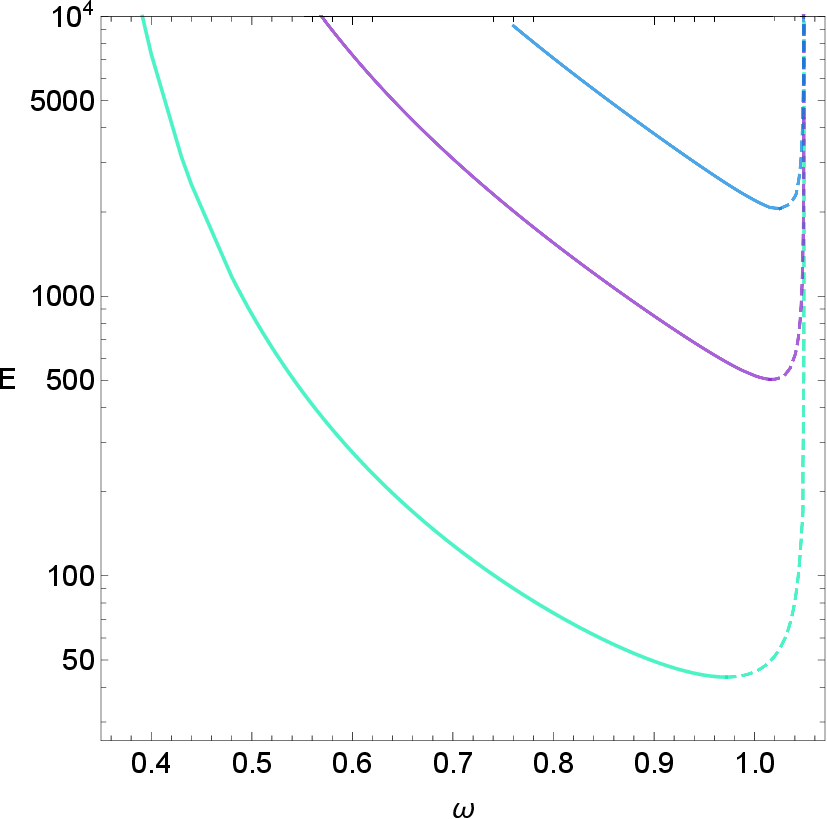}
\caption{The field $\phi(r)$ is shown as a function of the radial coordinate for the fundamental mode (green curves), the first (purple curves) and the second (blue curves) radial excited mode for $\omega=0.7$. We also show the evolution of the energy as a function of the frequency. The dashed region corresponds to the unstable solutions according to the fission stability criteria.}
\label{Fig:standardbis}
\end{figure}

Q-balls have also excited states that correspond to solutions with nodes but with the same limit at infinity, namely $\phi(\infty)=0$. In Fig. \ref{Fig:standardbis}, we show the first and second excited modes for a given frequency $\omega$. To fulfill the boundary conditions, for excited states, the initial conditions must be extremely fine-tuned. The excited states have as expected larger energy but also charge. We found that the frequency corresponding to $dE/d\omega=0$ becomes larger with the number of nodes. For example, for the fundamental mode, we have a minimal energy for $\omega=0.972$, while $\omega=1.015$ for the first excited mode and $\omega=1.025$ for the second excited mode. 

%\onecolumngrid
%\twocolumngrid

\begin{figure}
\centering
\includegraphics[scale=0.47]{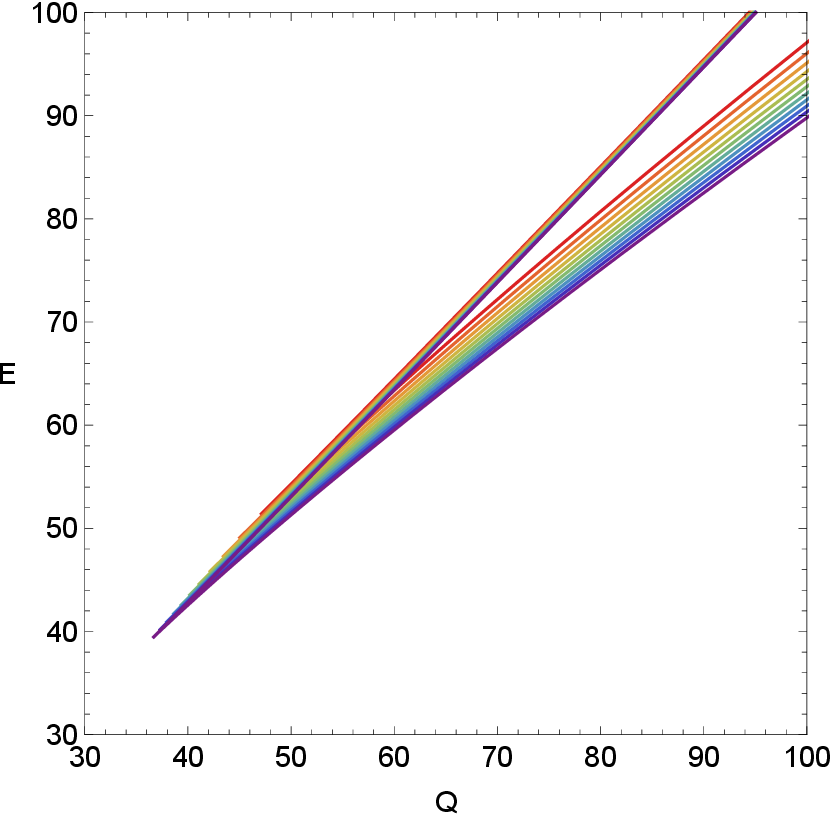}
\caption{The energy is shown as a function of charge for different values of the parameter $\alpha$ which runs from $\alpha=-0.5$ in red to $\alpha=0.5$ in purple with an incrementation of $0.1$}
\label{Fig:EQ}
\end{figure}

All these solutions are easily generalized to K-field theories. We will consider the simplest model where the action is modified by a single parameter, $K=X+\alpha X^2-V(|\Phi|)$ where $\alpha$ is the new parameter of the model\footnote{We assume our model corresponds to the low energy effective field theory where a small-$X$ expansion is possible and therefore terms $X^n$ with $n\geq 3$ are negligible. This is the complex analogous of \cite{Adams:2006sv} where $\alpha^{-1/4}$ is a cut off scale.}. Generically, we found that the structure of the solutions will not change. Q-balls exist for a certain range of frequency which depends on the parameter $\alpha$. We see from Fig. \ref{Fig:EQ} that for a given frequency, the Q-ball lowers its energy for large positive values of the parameter $\alpha$, because the radius decreases. 
%In the thin wall limit, we can consider that the field is almost always constant, therefore all derivatives can be neglected. Under that approximation, the K-field modifies only the potential. In our case, the self-interaction term $b\phi^4$ is modified into $(b+\alpha \omega^4)\phi^4$, lowering the potential and therefore the energy for positive values of $\alpha$. 
Notice that the critical value, $(E'(\omega)=0)$, of the energy and charge is also lowered for larger values of $\alpha$. Therefore, for a given frequency, the modified model with $\alpha>0$ produces Q-balls with lower charge and energy. The modification by the K-field allows one to build Q-balls with small charge and energy or on the contrary with larger energy and charge. Finally, we found that for all values of the parameter $\alpha$, in the limit of $\omega\rightarrow \omega_{\text{max}}$, or the thick-wall limit, we have the scaling solution $E=\omega Q^{\gamma}$ with $\gamma=1\pm10^{-4}$. This expression generalizes results found in \cite{Tsumagari:2008bv}.

\begin{figure}
\centering
\includegraphics[scale=0.61]{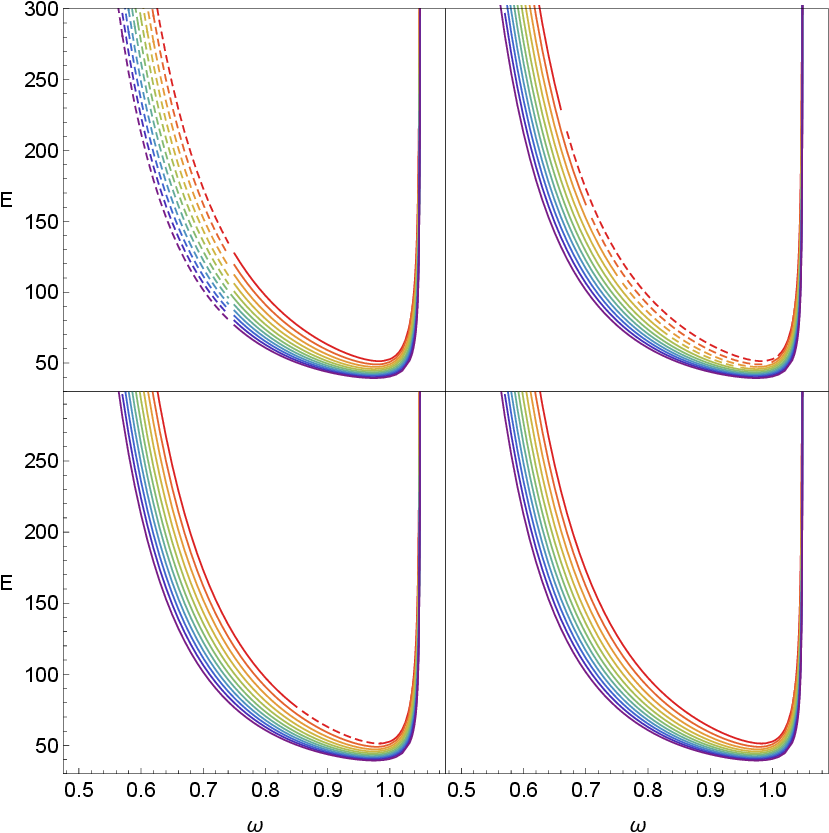}
\caption{Energy versus frequency for K-field model with $\alpha$ running from $-0.5$ (in red) to $+0.5$ (in purple) with a step of $0.1$. For each panel, we have represented in dotted lines the regime where some energy condition is violated. From top left to bottom right, we show the violation of the SEC, DEC, WEC, NEC.}
\label{Fig:EC}
\end{figure}

In Fig. \ref{Fig:EC}, we show the energy versus the frequency for different values of $\alpha$ but with the information on the violation of the energy conditions. We see that NEC is never violated. This condition corresponds to $1+2\alpha (\omega^2\phi^2-\phi'^2)>0$. It could be violated for very negative values of $\alpha$, but the construction of Q-balls for $\alpha<-0.5$ becomes very challenging and often impossible. In general, the larger and positive $\alpha$, the lower the probability to violate an energy condition, except for the SEC which is violated for any $\alpha$.

%\begin{figure}[H]
%\centering
%\includegraphics[scale=0.4]{stability.pdf}
%\caption{Using catastrophe theory which is similar than fission criteria, we represent the stability region.}
%\label{Fig:catastrophe}
%\end{figure}   

\section{Perturbations}

To study the mechanical stability, we decompose our field as 
\begin{align}
    \Phi(t,r,\theta,\varphi)=\phi(r)e^{i\omega t}+\sum_{\ell,m}\delta\Phi_{lm}(t,r) e^{i\omega t} Y_\ell^m(\theta,\varphi)\nonumber
\end{align}
where $\phi(r)$ is the background scalar field studied in the previous sections, $\delta \Phi_{\ell m}$ is the scalar field perturbation, $e^{i\omega t}$ in the second term is included for convenience and $Y_\ell^m$ are spherical harmonics. Because of the symmetries of the Q-balls, the perturbations will be independent of the azimuthal number $m$ and therefore the spherical harmonics reduce to Legendre polynomials. We will fix $m=0$. Notice that the different modes, $\ell$, do not couple and therefore we will omit this index. At second order in perturbations, and after integrating over the angle variables, the action reduces to 

\begin{align}
S&=\int {\rm d}t {\rm d}r \Bigl[
r^2 K_{,X} \dot\Psi_1^2-r^2(K_{,X}-2\phi'^2K_{,XX}) \Psi_1'^2
\nonumber\\
&+r^2 (K_{,X}+2\omega^2\phi^2 K_{,XX}) \dot\Psi_2^2-r^2 K_{,X} \Psi_2'^2
\nonumber\\
&-2\omega r^2 \phi \phi' K_{,XX}\Bigl(\dot\Psi_1 \Psi_2'+\Psi_1'\dot\Psi_2\Bigr)+A\Bigl(\dot\Psi_1\Psi_2-\Psi_1\dot\Psi_2\Bigr)\nonumber\\
&-M_1^2\Psi_1^2 -M_2^2\Psi_2^2\Bigr]
\label{eq:action}
\end{align}
where we have decomposed the perturbation into its real and imaginary parts, $\delta\Phi=\Psi_1+i \Psi_2$, and
\begin{align}
    A &= -2\omega r^2 \frac{{\rm d}}{\rm d(\phi^2)}\Bigl(\phi^2 K_{,X}\Bigr)-\omega \frac{{\rm d} }{{\rm d}r} \Bigl(r^2\phi\phi' K_{,XX}\Bigr)\label{eqA}\nonumber\\
    M_1^2 &=\lambda K_{,X} - \frac{r^2}{2} K_{,\phi\phi}
    -\frac{{\rm d}}{{\rm d} r}\Bigl(r^2\phi' K_{,X\phi}\Bigr)\nonumber\\
    M_2^2 &=\lambda K_{,X}- r^2 \Bigl(K_{,\phi^2}+\omega^2K_{,X}\Bigr)\nonumber\\
    \lambda &= \ell (\ell+1)
\end{align}
From this action, we obtain the two coupled equations for linear perturbations
\begin{align}
     & -K_{,X}\ddot\Psi_1+(K_{,X}-2\phi'^2K_{,XX})\Psi_1'' +2\omega \phi\phi' K_{,XX}\dot\Psi_2'\nonumber\\
    & \qquad +F_1(r,\Psi_1,\Psi_2,\Psi_1',\dot \Psi_2)=0 \label{eq:pert1}\\
    &-(K_{,X}+2\omega^2\phi^2 K_{,XX})\ddot\Psi_2+K_{,X}\Psi_2''+2\omega \phi\phi' K_{,XX}\dot\Psi_1'\nonumber\\
    & \qquad +F_2(r,\Psi_1,\Psi_2,\Psi_2',\dot \Psi_1)=0
    \label{eq:pert2}
\end{align}
with $F_1$ and $F_2$ some functions defined by the perturbations and their first derivative. 

These equations form a set of two coupled differential equations representing the evolution of the perturbations in an effective metric. Indeed, if we consider, e.g. Eq.(\ref{eq:pert1}), and in the absence of coupling between $\Psi_1$ and $\Psi_2$, i.e., $\omega=0$, the equation would reduce to the generic form $h^{\mu\nu}\nabla_{\mu\nu}\Psi_1+\cdots=0$, with $h^{00}=-K_{,X}$ and $h^{11}=K_{,X}-2\phi'^2 K_{,XX}$, from which we obtain the stability conditions, i.e., a Lorentzian effective metric $h^{00}<0$ and $h^{11}>0$. These conditions are equivalent to the Hamiltonian of field perturbations to be positive definite; indeed, as seen from Eq.(\ref{eq:action}), the Lagrangian (of $\Psi_1$ in the case of $\omega=0$) reduces to $L=r^2 (-h^{00}\dot\Psi^2-h^{11}\Psi'^2)$ and therefore to a Hamiltonian $H=r^2(-h^{00}\dot\Psi^2+h^{11}\Psi'^2)$. The Hamiltonian is bounded from below \cite{Arkani-Hamed:2003pdi,Armendariz-Picon:2005oog} if we satisfy the conditions for an effective Lorentzian metric 
\begin{align}
    &K_{,X}>0 \nonumber\\
    &K_{,X}-2\phi'^2 K_{,XX}>0 \Leftrightarrow K_{,X}+2X K_{,XX}>0 ~~(\omega=0)\nonumber
\end{align}
But as nicely stated in \cite{Babichev:2018uiw}, one should be careful, because Hamiltonian unboundedness is not always equivalent to stability. A Hamiltonian can be unbounded only because of our set of variables chosen. Therefore, stability should be imposed only from the existence of a future causal cone defined by the effective metric. In conclusion, to study stability, we need to ensure that the problem is well-posed. For that, we will derive the conditions of weak and strong hyperbolicity. Broadly speaking, the weak hyperbolicity condition forbids any solution to grow exponentially in time while the strong hyperbolicity condition imposes a stronger bound than the exponential growth and therefore is equivalent to local well-posedness of the Cauchy problem. In the case of a strong hyperbolic system, $F_1$ and $F_2$ will not be relevant while they could change the behavior of the system if weakly hyperbolic. We define the vector $u=(\Psi_1,\Psi_2)^T$ and the system (\ref{eq:pert1}) and (\ref{eq:pert2}) becomes
\begin{align}
u_{,tt}=A u''+B  u_{,t}'+\cdots 
\end{align}
where $\cdots $ indicates the lowest derivative terms, and
\begin{align}
    A_{11} &=\frac{K_{,X}-2\phi'^2 K_{,XX}}{K_{,X}}\\
    A_{22} &=\frac{K_{,X}}{K_{,X}+2\omega^2 \phi^2 K_{,XX}}\\
    B_{12} &=2\omega \phi\phi' \frac{K_{,XX}}{K_{,X}}\\
    B_{21} &=2\omega \phi\phi' \frac{K_{,XX}}{K_{,X}+2\omega^2\phi^2K_{,XX}}
\end{align}
while other elements of the matrices are zero. We consider wave solutions $u(t,r)=e^{i k r} \hat u(t,k)$ and obtain 
\begin{align}
    \hat u_{,tt}=-k^2 A\hat u+ik B \hat u_{,t}+\cdots 
\end{align}
This system can be reduced to first order by defining the variable $\hat v=\hat u_{,t}/(i|k|)$
\begin{align}
    \begin{pmatrix}
        \hat v\\
        \hat u
    \end{pmatrix}_{,t}=i|k| \hat P\begin{pmatrix}
        \hat v\\
        \hat u
    \end{pmatrix}
\end{align}
with
\begin{align}
    \hat P =\begin{pmatrix}
        0 & \frac{k}{|k|}B_{12} & A_{11}  & 0\\
        \frac{k}{|k|}B_{21} & 0 & 0 & A_{22}\\
        1 & 0 & 0 & 0\\
        0 & 1 & 0 & 0
    \end{pmatrix}
\end{align}
The well-posedness of this system is reduced to the analysis of  the matrix $\hat P$ (see e.g. \cite{Kreiss:2001cu}). If, for all $k$, the eigenvalues of $\hat P$ are real, the system is weakly hyperbolic. The eigenvalues are
\begin{align}
    \Bigl\{\pm 1,\pm \sqrt{\frac{K_{,X}-2\phi'^2 K_{,XX}}{K_{,X}+2\omega^2\phi^2 K_{,XX}}}\Bigr\}
\end{align}
Therefore, we conclude that, if $\frac{K_{,X}-2\phi'^2 K_{,XX}}{K_{,X}+2\omega^2\phi^2 K_{,XX}}\geq 0$, the system is weakly hyperbolic. Additionally, when
\begin{align}
    \frac{K_{,X}-2\phi'^2 K_{,XX}}{K_{,X}+2\omega^2\phi^2 K_{,XX}} > 0
\end{align}
the system is strongly hyperbolic because the eigenvectors form a complete set. In that case, the two perturbations propagate at the speed
\begin{align}
    c_1 =1\,,\qquad 
    c_2 = \sqrt{\frac{K_{,X}-2\phi'^2 K_{,XX}}{K_{,X}+2\omega^2\phi^2 K_{,XX}}}
\end{align}
As we have shown in Sec. \ref{section:potential}, we consider the condition $K_{,X}-2\phi'^2K_{,XX}>0$ which implies $K_{,X}+2\omega^2 \phi^2K_{,XX}>0$. Summing these two conditions, we find a weaker condition, viz. $K_{,X}>0$ and $K_{,X}+X K_{,XX}>0$. Notice that for a real scalar field $(\omega=0)$, the condition $K_{,X}+2\omega^2 \phi^2K_{,XX}>0$ reduces to $K_{,X}>0$ along with the condition $K_{,X}-2\phi'^2K_{,XX}>0$ ($K_{,X}+2X K_{,XX}>0$), and they correspond to the stability conditions obtained in \cite{Armendariz-Picon:2005oog}.

\begin{figure}
\centering
\includegraphics[scale=0.5]{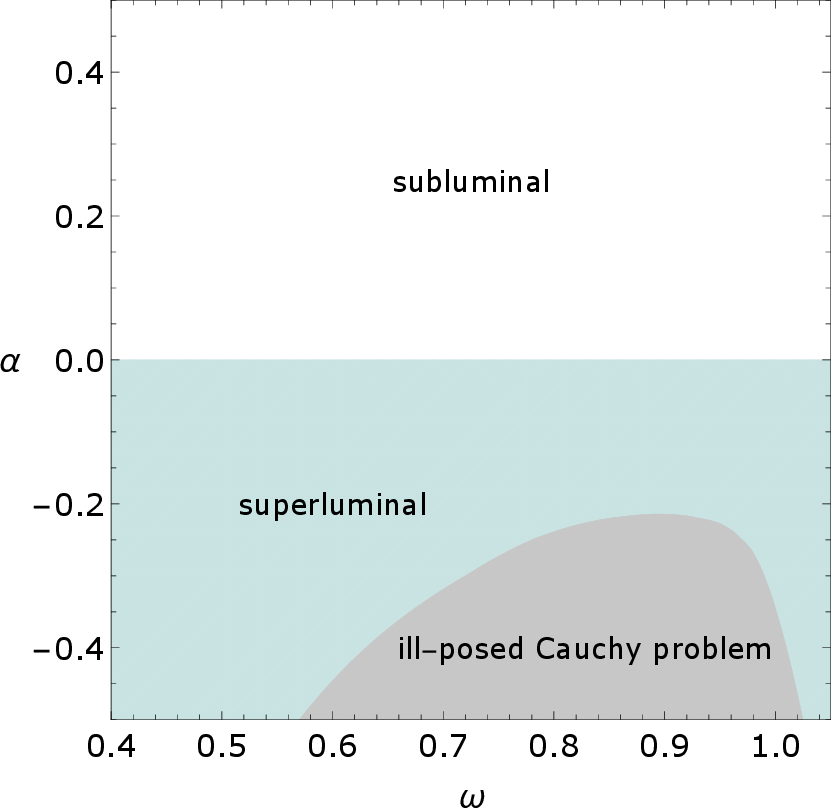}
\caption{In gray, the region of parameter space $(\omega,\alpha)$ where the Cauchy problem is not well-posed and in cyan the region of superluminal propagation.}
\label{Fig:strong}
\end{figure}

Notice that the conditions of well-posedness of the system are independent of the energy conditions derived previously (\ref{SEC}), (\ref{DEC}), (\ref{WEC}) and \ref{NEC}). In Fig. \ref{Fig:strong}, and for the model $K=X+\alpha X^2-V(\phi)$, we have found that for a certain range of the parameters $(\omega,\alpha)$, the Cauchy problem is not well-posed which never corresponds to $\alpha>0$. Also we found that for any $\alpha<0$, the perturbations are superluminal in some region of space. Even if the classical theory is well-posed, the superluminal propagation of the perturbations could be an obstacle to a quantum version of the theory. For example, requiring UV completion for K-essence (real scalar field analog to the case studied in this paper) imposes subluminal propagation \cite{Melville:2019wyy}. A similar situation should be expected in our case \cite{Adams:2006sv}. Even if not equivalent, we found numerically, for all parameters $(\omega,\alpha)$ of Fig. \ref{Fig:strong}, that a system which violates the weak energy condition does not have a well-posed Cauchy problem. The converse is not true.

Restricting our analysis to the cases where the Cauchy problem is well-posed, we can study the mechanical stability of our solutions. For that, we assume the following form for the perturbation:
\begin{align}
    \delta\Phi(t,r)=\frac{\eta(r)}{r^n} e^{i\rho t}+\frac{\chi^*(r)}{r^n} e^{-i\rho^* t}
    \label{eq:pertu}
\end{align}
The system (\ref{eq:pert1}) and (\ref{eq:pert2}) reduces to two ordinary coupled differential equations for $\eta(r)$ and $\chi(r)$. We have included a factor $r^n$ for numerical stability. In general, $n=\ell$ provides faster numerical results. In the canonical case where $K_{,X}=1$, the stability analyses shows that any instability corresponds to $\rho=-\rho^*$ \cite{Panin:2016ooo} which implies the condition $\frac{dQ}{d\omega}<0$. We could not extend this analysis to K-field theories and therefore we will study the perturbations by numerical means. For that, our system can be written as four first order differential equations for the variable $\Psi\equiv(\eta,\chi,\eta',\chi')^T$, $\Psi'=B\Psi$ where the matrix $B$ is given in Appendix \ref{pert}. Considering the conditions at $r=0$ on the scalar field, $\phi'=0$, it is easy to show that perturbations behave as
\begin{align}
    \eta(r\simeq 0) &= c_0 r^{\ell+n}\\
    \chi(r\simeq 0) &= c_1 r^{\ell+n}
\end{align}
which implies
\begin{align}
    \Psi(0)=c_0 r^{\ell+n-1}\begin{pmatrix}
        r\\
        0\\
        \ell+n\\
        0
    \end{pmatrix}+c_1r^{\ell+n-1}\begin{pmatrix}
        0\\
        r\\
        0\\
        \ell+n
    \end{pmatrix}
\end{align}
Therefore, we can perform two numerical integrations from $r=0$ with initial conditions $\eta=r^{\ell+n}, \chi=0$ and $\eta=0, \chi=r^{\ell+n}$ respectively. The general solution will be a linear combination of these two solutions with coefficients $(c_0,c_1)$. Similarly, we perform an integration from infinity to $r=0$. We have also a system with two free parameters $(c_3,c_4)$. We can integrate it from a large radius with initial conditions 
\begin{align}
    \eta = \frac{e^{-r \sqrt{-\frac{K_{,\phi^2}(0,0)}{K_{,X}(0,0)}-(\rho+\omega)^2}}}{r^{1-n}}\,,~~\chi=0
\end{align}
or
\begin{align}
    \chi = \frac{e^{-r \sqrt{-\frac{K_{,\phi^2}(0,0)}{K_{,X}(0,0)}-(\rho-\omega)^2}}}{r^{1-n}}\,,~~\eta=0
\end{align}
Having the solution integrated from both boundaries with four free parameters $(c_1,c_2,c_3,c_4)$, we can match them at a given radius, using the four continuity conditions of $(\eta,\chi,\eta',\chi')$. Notice that, because our system is linear, we can always fix one of the parameters, e.g. $c_1=1$. Therefore, we end with a system of four conditions and three parameters, the fourth parameter will determine the value of $\rho$. In conclusion, only a certain number of discrete values of $\rho$ can solve our problem. 

\begin{figure}
\centering
\includegraphics[scale=0.75]{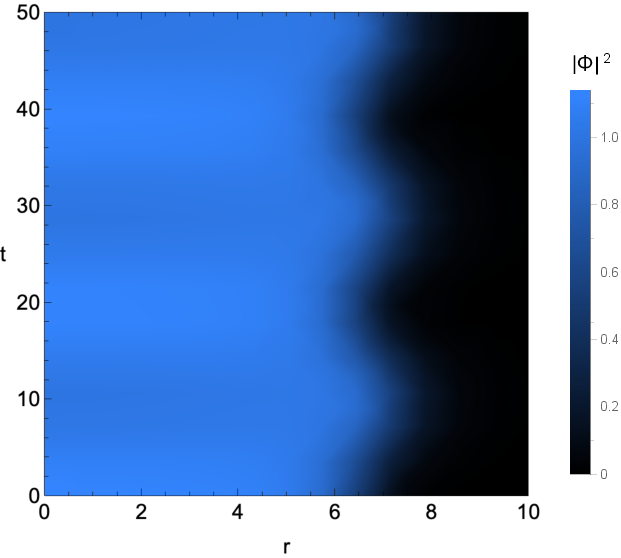}
\includegraphics[scale=0.75]{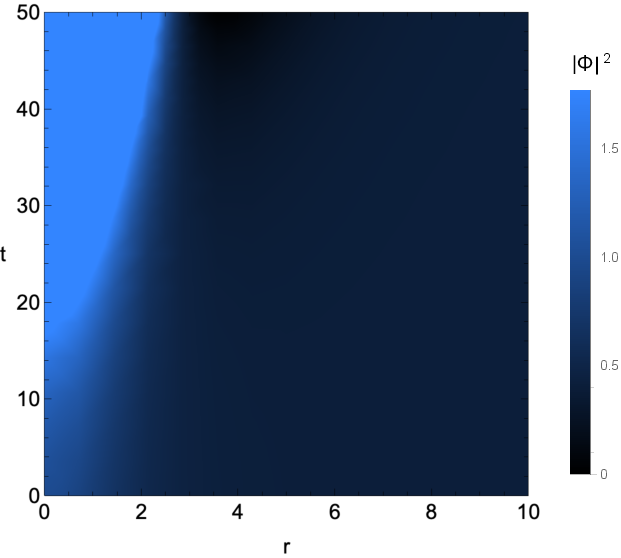}
\caption{Spacetime diagram of $|\Phi|^2$. The upper diagram shows the stability of the background solution with $\omega=0.5$ and the lower case shows an unstable solution for $\omega=1$. For both solutions, we have considered $\alpha=0$.}
\label{Fig:stability}
\end{figure}

In Fig. (\ref{Fig:stability}), we show $|\phi+\delta\Phi|^2$, for $\omega=(0.5,1)$ and $\alpha=0$. For each case, we have found the parameter $\rho$ and using Eq.(\ref{eq:pertu}), we obtain the time and space dependence of the solution. In the case, where $\omega=0.5$, the radius of the Q-ball is oscillating, and $\rho$ is real. The energy of this solution is constant in time, while for $\omega=1$, the energy grows exponentially as well as the radius of the Q-ball. The solution is unstable and $\rho$ is purely imaginary.  

Therefore, the strategy is simple, for each Q-ball, we search in the complex plane for values of the $\rho$ solution to our previous problem. 

For the excited states, all frequencies $\omega$ were unstable. But for various frequencies, the unstable modes were not always purely imaginary but also with a nonzero real part.

For the fundamental solution, Fig. \ref{Fig:stability} shows two cases where $\alpha=0$ and $\omega=(0.5,1)$. The first frequency corresponds to a stable solution for which we see an oscillation of the radius of the Q-ball while the energy remains perfectly constant in time. The second case, corresponds to an unstable solution for which the radius increases and the energy grows exponentially. 

Generically, we found that the stability region corresponds to $dQ/d\omega<0$ for all $\omega$, generalizing results that were known in the canonical case. In the unstable region, the timescale of the instability is of the order $1/\text{Im}(\rho)$. We found that Im$(\rho)$ and therefore the timescale of the instability depends on the mode $\ell$. For example, for $\alpha=0$, Im$(\rho)$ is of the order $10^{-1}$ for $\ell=0$ and of the order $10^{-4}$ for $\ell=1$. Therefore, we will focus mainly on the spherical mode of perturbations $\ell=0$.

\begin{figure}
\centering
\includegraphics[scale=.85]{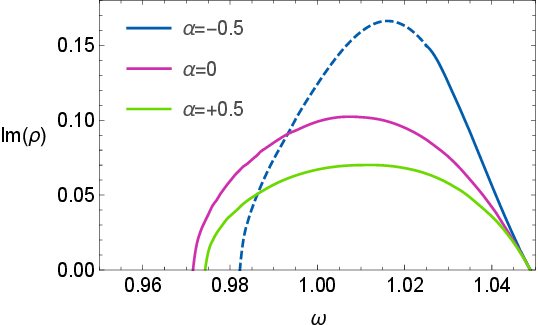}
\caption{Existence of Im$(\rho)$ as a function of $\omega$ for $\alpha=(-0.5,0,+0.5)$. The existence of such a mode implies an instability of the background solution. The dotted line corresponds to unstable modes but in a region where the Cauchy problem is not well-posed and therefore should be excluded from the analysis.}
\label{Fig:frequency}
\end{figure}

In Fig. \ref{Fig:frequency}, we show the unstable modes for three values of $\alpha$. For each $\alpha$, the instability starts when $dQ/d\omega=0$. We notice also that even if for a given frequency, such as $\omega=1.03$, the Q-ball is unstable for all values of the parameter $\alpha$, the instability is slower to develop (lower value of Im$(\rho)$), for larger positive values of $\alpha$, which is consistent with the previous section where we found that the energy is lowered.

\begin{figure}
\centering
\includegraphics[scale=.85]{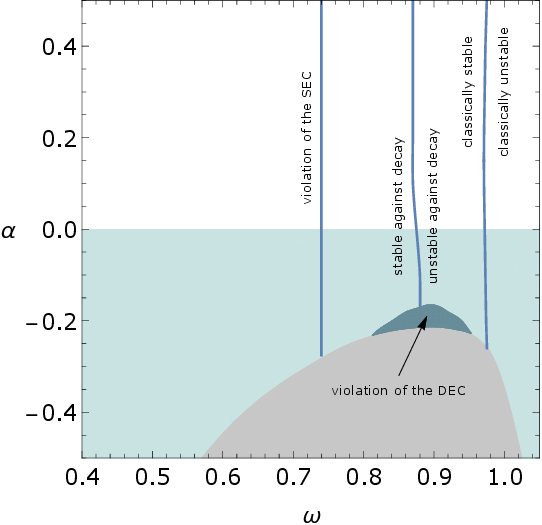}
\caption{Space of parameters $(\omega,\alpha)$ within the region where the Cauchy problem is well-posed. We have represented regions of quantum stability (against fission) and classical stability as well as regions where the energy conditions such as the SEC and DEC are violated. We have kept the cyan and white colors for, respectively, superluminal and subluminal  propagation.}
\label{Fig:summary}
\end{figure}

In Fig. \ref{Fig:summary}, we summarize the various stability conditions. The quantum stability condition, namely the stability against fission is, as expected, stronger than the classical stability condition.  We have also represented regions where the energy conditions are violated. The NEC is never violated in the region of analysis of the model while the WEC is violated only in the region where the Cauchy problem is not well-posed. The violation of the SEC and the DEC are totally independent of the stability conditions.

\section{Conclusion}
In this work we studied Q-balls in noncanonical scalar field theory. We derived the general equations of existence and stability for these theories. We found that the stability against fission and the linear mechanical stability are equivalent and reduce to $Q'(\omega)<0$ (see Table \ref{Table1}). On the other hand, the condition for decay into free particles is stronger. 

\begin{table}[htbp]
\footnotesize
\begin{tabular}{ |p{1.6cm}||p{1.6cm}||p{4.8cm}|}
 \hline
 \multicolumn{3}{|c|}{Stability conditions} \\
 \hline\xrowht[()]{10pt}
 Fission & $\frac{dQ}{d\omega}<0$ & Proved for K-field theories\\
 \hline\xrowht[()]{10pt}
 Decay   & $E<M Q$ & Generic condition (for any theory)\\
 \hline\xrowht[()]{10pt}
 Classical &  $\frac{dQ}{d\omega}<0$ & Shown numerically to be the same condition for $K=X+\alpha X^2-V(|\Phi|^2)$\\
 \hline
\end{tabular}
\caption{\small Summary of the three stability conditions studied in this paper and extended to K-field theories.}
\label{Table1}
\end{table}

We found that perturbations have a well-posed Cauchy problem if $\frac{K_{,X}-2\phi'^2 K_{,XX}}{K_{,X}+2\omega^2\phi^2 K_{,XX}} > 0$. When the perturbations are strongly hyperbolic, we found that perturbations are superluminal or subluminal. In the particular case, $K=X+\alpha X^2-V(|\Phi|^2)$, perturbations are subluminal and luminal for $\alpha>0$ while they are superluminal and luminal for $\alpha<0$. 
We found that a Q-ball with $\alpha>0$ lowers its energy for larger values of $\alpha$. Even in the unstable region, the timescale of this instability becomes larger and therefore more stable. The frequency at which Q-balls become unstable increases with $\alpha$. It would be interesting to find models for which all Q-balls are stable irrespectively of their frequency.

Finally, we have studied the different energy conditions such as the SEC, DEC, WEC, NEC. We found that NEC is never violated and none of these conditions can be related to mechanical stability.

\section*{Acknowledgements}
The work of A.F. is supported by ANID/CONICYT No. 21171262 while R.G. is supported by ANID FONDECYT Regular No. 1220965.

\appendix*
\section{Perturbation equations}
\label{pert}

The matrix of the system $\Psi'=B\Psi$ can be decomposed as $B=B^{(0)}+\rho B^{(1)}+\rho^2 B^{(2)}$, where
\begin{align}
    B^{(0)} &=\begin{pmatrix}
        0 & 0 & 1 & 0\\
        0 & 0 & 0 & 1\\
        B^{(0)}_{31} & B^{(0)}_{32} & B^{(0)}_{33} & B^{(0)}_{34}\\
        B^{(0)}_{32} & B^{(0)}_{31} & B^{(0)}_{34} & B^{(0)}_{33}
    \end{pmatrix}\\
    B^{(1)} &=\begin{pmatrix}
        0 & 0 & 0 & 0\\
        0 & 0 & 0 & 0\\
        B^{(1)}_{31} & B^{(1)}_{32} & B^{(1)}_{33} & B^{(1)}_{34}\\
        -B^{(1)}_{32} & -B^{(1)}_{31} & -B^{(1)}_{34} & -B^{(1)}_{33}
    \end{pmatrix}\\
    B^{(2)} &=\begin{pmatrix}
        0 & 0 & 0 & 0\\
        0 & 0 & 0 & 0\\
        B^{(2)}_{31} & B^{(2)}_{32} & 0 & 0\\
        B^{(1)}_{32} & B^{(1)}_{31} & 0 & 0
    \end{pmatrix}
\end{align}
\vspace{-0.4cm}

with
\begin{align}
    B^{(0)}_{31} &= \frac{1}{2r^2}\Bigl(\frac{M_1^2}{K_{,X}-2\phi'^2K_{,XX}}+\frac{M_2^2}{K_{,X}}\Bigr)\\
    B^{(0)}_{32} &= \frac{1}{2r^2}\Bigl(\frac{M_1^2}{K_{,X}-2\phi'^2K_{,XX}}-\frac{M_2^2}{K_{,X}}\Bigr)\\
    B^{(0)}_{33} &= -\frac{2}{r}-\frac{1}{2}\frac{{\rm d}}{{\rm d} r}\log\Bigl(K_{,X}(K_{,X}-2\phi'^2 K_{,XX})\Bigr)\\
    B^{(0)}_{34} &=\frac{1}{2}\frac{{\rm d}}{{\rm d} r}\log\frac{K_{,X}}{K_{,X}-2\phi'^2 K_{,XX}}
\end{align}
\begin{align}
    B^{(1)}_{31} &= A\frac{K_{,X}-\phi'^2K_{,XX}}{r^2 K_{,X} (K_{,X}-2\phi'^2 K_{,XX})}\nonumber\\
    &\quad -\frac{\omega \Bigl(r^4 \phi\phi'^3 K_{,XX}^2\Bigr)'}{2 r^4 K_{,X}(K_{,X}-2\phi'^2 K_{,XX})}\nonumber\\
    &\quad -\frac{\omega \phi'^2(\phi'^2-\phi\phi'')K_{,XX}^2}{2 K_{,X}(K_{,X}-2\phi'^2 K_{,XX})}\\
    B^{(1)}_{32} &= \frac{\omega \Bigl(\phi\phi'K_{,XX}\Bigr)'}{K_{,X}-2\phi'^2 K_{,XX}}+\frac{2\omega \phi\phi' K_{,X}K_{,XX}}{rK_{,X}(K_{,X}-2\phi'^2 K_{,XX})}\nonumber\\
    & \hspace{-0.5cm}+2\omega r \phi'^2K_{,XX}\frac{K_{,X}+\omega^2\phi^2K_{,XX}+\phi^2 K_{,X\phi^2}}{rK_{,X}(K_{,X}-2\phi'^2 K_{,XX})}
\end{align}
\begin{align}
    B^{(1)}_{33} &= -\frac{2\omega\phi\phi'^3 K_{,XX}^2}{K_{,X}(K_{,X}-2\phi'^2 K_{,XX})}\\
    B^{(1)}_{34} &= 2 \omega\phi\phi' K_{,XX}\frac{K_{,X}-\phi'^2K_{,XX}}{K_{,X} (K_{,X}-2\phi'^2 K_{,XX})}\\
    B^{(2)}_{31} &= \frac{2\omega^2\phi^2\phi'^2K_{,XX}^2-K_{,X}^2-XK_{,X}K_{,XX}}{K_{,X}(K_{,X}-2\phi'^2K_{,XX})}\\
    B^{(2)}_{32} &= K_{,XX}\Bigl(\frac{\omega^2\phi^2}{K_{,X}}-\frac{\phi'^2}{K_{,X}-2\phi'^2 K_{,XX}}\Bigr)
\end{align}
and $(A,M_1^2,M_2^2)$ are defined by Eq.(\ref{eqA}). These equations are given in the case of $n=0$.


\begin{thebibliography}{99}

%\cite{Rosen:1968mfz}
\bibitem{Rosen:1968mfz}
G.~Rosen,
%``Particlelike Solutions to Nonlinear Complex Scalar Field Theories with Positive-Definite Energy Densities,''
J. Math. Phys. \textbf{9} (1968), 996
%doi:10.1063/1.1664693
%166 citations counted in INSPIRE as of 22 Aug 2022

%\cite{Coleman:1985ki}
\bibitem{Coleman:1985ki}
S.~R.~Coleman,
%``Q-balls,''
Nucl. Phys. B \textbf{262} (1985) no.2, 263
%doi:10.1016/0550-3213(86)90520-1
%915 citations counted in INSPIRE as of 14 May 2022

%\cite{Freese:1990rb}
\bibitem{Freese:1990rb}
K.~Freese, J.~A.~Frieman and A.~V.~Olinto,
%``Natural inflation with pseudo - Nambu-Goldstone bosons,''
Phys. Rev. Lett. \textbf{65} (1990), 3233-3236
%doi:10.1103/PhysRevLett.65.3233
%1193 citations counted in INSPIRE as of 17 Aug 2022

%\cite{Adams:1992bn}
\bibitem{Adams:1992bn}
F.~C.~Adams, J.~R.~Bond, K.~Freese, J.~A.~Frieman and A.~V.~Olinto,
%``Natural inflation: Particle physics models, power law spectra for large scale structure, and constraints from COBE,''
Phys. Rev. D \textbf{47} (1993), 426-455
%doi:10.1103/PhysRevD.47.426
[arXiv:hep-ph/9207245 [hep-ph]].
%613 citations counted in INSPIRE as of 17 Aug 2022

%\cite{Kusenko:1997zq}
\bibitem{Kusenko:1997zq}
A.~Kusenko,
%``Solitons in the supersymmetric extensions of the standard model,''
Phys. Lett. B \textbf{405} (1997), 108
%doi:10.1016/S0370-2693(97)00584-4
[arXiv:hep-ph/9704273 [hep-ph]].
%294 citations counted in INSPIRE as of 23 Aug 2022

%\cite{Affleck:1984fy}
\bibitem{Affleck:1984fy}
I.~Affleck and M.~Dine,
%``A New Mechanism for Baryogenesis,''
Nucl. Phys. B \textbf{249} (1985), 361-380
%doi:10.1016/0550-3213(85)90021-5
%1332 citations counted in INSPIRE as of 22 Aug 2022

%\cite{Dine:1995kz}
\bibitem{Dine:1995kz}
M.~Dine, L.~Randall and S.~D.~Thomas,
%``Baryogenesis from flat directions of the supersymmetric standard model,''
Nucl. Phys. B \textbf{458} (1996), 291-326
%doi:10.1016/0550-3213(95)00538-2
[arXiv:hep-ph/9507453 [hep-ph]].
%757 citations counted in INSPIRE as of 22 Aug 2022

%\cite{Enqvist:2003gh}
\bibitem{Enqvist:2003gh}
K.~Enqvist and A.~Mazumdar,
%``Cosmological consequences of MSSM flat directions,''
Phys. Rept. \textbf{380} (2003), 99-234
%doi:10.1016/S0370-1573(03)00119-4
[arXiv:hep-ph/0209244 [hep-ph]].
%319 citations counted in INSPIRE as of 22 Aug 2022

%\cite{Kusenko:1997si}
\bibitem{Kusenko:1997si}
A.~Kusenko and M.~E.~Shaposhnikov,
%``Supersymmetric Q balls as dark matter,''
Phys. Lett. B \textbf{418} (1998), 46-54
%doi:10.1016/S0370-2693(97)01375-0
[arXiv:hep-ph/9709492 [hep-ph]].
%624 citations counted in INSPIRE as of 22 Aug 2022

%\cite{Kusenko:2001vu}
\bibitem{Kusenko:2001vu}
A.~Kusenko and P.~J.~Steinhardt,
%``Q ball candidates for selfinteracting dark matter,''
Phys. Rev. Lett. \textbf{87} (2001), 141301
%doi:10.1103/PhysRevLett.87.141301
[arXiv:astro-ph/0106008 [astro-ph]].
%137 citations counted in INSPIRE as of 22 Aug 2022

%\cite{Friedberg:1976me}
\bibitem{Friedberg:1976me}
R.~Friedberg, T.~D.~Lee and A.~Sirlin,
%``A Class of Scalar-Field Soliton Solutions in Three Space Dimensions,''
Phys. Rev. D \textbf{13} (1976), 2739-2761
%doi:10.1103/PhysRevD.13.2739
%417 citations counted in INSPIRE as of 22 Aug 2022

%\cite{Lee:1991ax}
\bibitem{Lee:1991ax}
T.~D.~Lee and Y.~Pang,
%``Nontopological solitons,''
Phys. Rept. \textbf{221} (1992), 251-350
%doi:10.1016/0370-1573(92)90064-7
%467 citations counted in INSPIRE as of 22 Aug 2022

%\cite{Panin:2016ooo}
\bibitem{Panin:2016ooo}
A.~G.~Panin and M.~N.~Smolyakov,
%``Problem with classical stability of U(1) gauged Q-balls,''
Phys. Rev. D \textbf{95} (2017) no.6, 065006
%doi:10.1103/PhysRevD.95.065006
[arXiv:1612.00737 [hep-th]].
%24 citations counted in INSPIRE as of 22 Aug 2022

%\cite{Adams:2006sv}
\bibitem{Adams:2006sv}
A.~Adams, N.~Arkani-Hamed, S.~Dubovsky, A.~Nicolis and R.~Rattazzi,
%``Causality, analyticity and an IR obstruction to UV completion,''
JHEP \textbf{10} (2006), 014
%doi:10.1088/1126-6708/2006/10/014
[arXiv:hep-th/0602178 [hep-th]].
%850 citations counted in INSPIRE as of 27 Mar 2023

%\cite{Tsumagari:2008bv}
\bibitem{Tsumagari:2008bv}
M.~I.~Tsumagari, E.~J.~Copeland and P.~M.~Saffin,
%``Some stationary properties of a Q-ball in arbitrary space dimensions,''
Phys. Rev. D \textbf{78} (2008), 065021
%doi:10.1103/PhysRevD.78.065021
[arXiv:0805.3233 [hep-th]].
%55 citations counted in INSPIRE as of 22 Aug 2022

%\cite{Silverstein:2003hf}
\bibitem{Silverstein:2003hf}
E.~Silverstein and D.~Tong,
%``Scalar speed limits and cosmology: Acceleration from D-cceleration,''
Phys. Rev. D \textbf{70} (2004), 103505
%doi:10.1103/PhysRevD.70.103505
[arXiv:hep-th/0310221 [hep-th]].
%767 citations counted in INSPIRE as of 22 Aug 2022

%\cite{Goon:2011qf}
\bibitem{Goon:2011qf}
G.~Goon, K.~Hinterbichler and M.~Trodden,
%``Symmetries for Galileons and DBI scalars on curved space,''
JCAP \textbf{07} (2011), 017
%doi:10.1088/1475-7516/2011/07/017
[arXiv:1103.5745 [hep-th]].
%137 citations counted in INSPIRE as of 22 Aug 2022

%\cite{Sen:2002an}
\bibitem{Sen:2002an}
A.~Sen,
%``Field theory of tachyon matter,''
Mod. Phys. Lett. A \textbf{17} (2002), 1797-1804
%doi:10.1142/S0217732302008071
[arXiv:hep-th/0204143 [hep-th]].
%556 citations counted in INSPIRE as of 22 Aug 2022

%\cite{Kuniyasu:2016tse}
\bibitem{Kuniyasu:2016tse}
M.~Kuniyasu, N.~Sakai and K.~Shiraishi,
%``Q -balls in DBI type k field theory,''
Phys. Rev. D \textbf{94} (2016) no.11, 116001
%doi:10.1103/PhysRevD.94.116001
%1 citations counted in INSPIRE as of 23 Aug 2022

%\cite{Sakai:2007ft}
\bibitem{Sakai:2007ft}
N.~Sakai and M.~Sasaki,
%``Stability of Q-balls and Catastrophe,''
Prog. Theor. Phys. \textbf{119} (2008), 929-937
%doi:10.1143/PTP.119.929
[arXiv:0712.1450 [hep-ph]].
%32 citations counted in INSPIRE as of 22 Aug 2022

%\cite{Armendariz-Picon:1999hyi}
\bibitem{Armendariz-Picon:1999hyi}
C.~Armendariz-Picon, T.~Damour and V.~F.~Mukhanov,
%``k - inflation,''
Phys. Lett. B \textbf{458} (1999), 209-218
%doi:10.1016/S0370-2693(99)00603-6
[arXiv:hep-th/9904075 [hep-th]].
%1765 citations counted in INSPIRE as of 03 Jan 2023

%\cite{Armendariz-Picon:2000nqq}
\bibitem{Armendariz-Picon:2000nqq}
C.~Armendariz-Picon, V.~F.~Mukhanov and P.~J.~Steinhardt,
%``A Dynamical solution to the problem of a small cosmological constant and late time cosmic acceleration,''
Phys. Rev. Lett. \textbf{85} (2000), 4438-4441
%doi:10.1103/PhysRevLett.85.4438
[arXiv:astro-ph/0004134 [astro-ph]].
%1729 citations counted in INSPIRE as of 03 Jan 2023

\cite{VK}
\bibitem{VK}
N.~G.~Vakhitov and A.~A.~Kolokolov, 
Radiophysics and Quantum Electronics, 16(7), (1973) pp.783-789.

%\cite{Schunck:1992}
\bibitem{Schunck:1992}
F.E. Schunck, F.V. Kusmartsev, E.W. Mielke, in: R.A. d'Inverno (Ed.), Approaches to Numerical Relativity,
Cambridge Univ. Press, Cambridge, 1992, pp. 130–140.

%\cite{Volkov:2002aj}
\bibitem{Volkov:2002aj}
M.~S.~Volkov and E.~Wohnert,
%``Spinning Q balls,''
Phys. Rev. D \textbf{66} (2002), 085003
%doi:10.1103/PhysRevD.66.085003
%[arXiv:hep-th/0205157 [hep-th]].
%104 citations counted in INSPIRE as of 15 May 2022

%\cite{Arkani-Hamed:2003pdi}
\bibitem{Arkani-Hamed:2003pdi}
N.~Arkani-Hamed, H.~C.~Cheng, M.~A.~Luty and S.~Mukohyama,
%``Ghost condensation and a consistent infrared modification of gravity,''
JHEP \textbf{05} (2004), 074
%doi:10.1088/1126-6708/2004/05/074
[arXiv:hep-th/0312099 [hep-th]].
%1017 citations counted in INSPIRE as of 26 Mar 2023

%\cite{Armendariz-Picon:2005oog}
\bibitem{Armendariz-Picon:2005oog}
C.~Armendariz-Picon and E.~A.~Lim,
%``Haloes of k-essence,''
JCAP \textbf{08} (2005), 007
%doi:10.1088/1475-7516/2005/08/007
[arXiv:astro-ph/0505207 [astro-ph]].
%153 citations counted in INSPIRE as of 26 Mar 2023

%\cite{Babichev:2018uiw}
\bibitem{Babichev:2018uiw}
E.~Babichev, C.~Charmousis, G.~Esposito-Far\`ese and A.~Leh\'ebel,
%``Hamiltonian unboundedness vs stability with an application to Horndeski theory,''
Phys. Rev. D \textbf{98} (2018) no.10, 104050
%doi:10.1103/PhysRevD.98.104050
[arXiv:1803.11444 [gr-qc]].
%53 citations counted in INSPIRE as of 26 Mar 2023


%\cite{Kreiss:2001cu}
\bibitem{Kreiss:2001cu}
H.~O.~Kreiss and O.~E.~Ortiz,
%``Some mathematical and numerical questions connected with first and second order time dependent systems of partial differential equations,''
Lect. Notes Phys. \textbf{604} (2002), 359
[arXiv:gr-qc/0106085 [gr-qc]].
%30 citations counted in INSPIRE as of 16 Dec 2022

%\cite{Melville:2019wyy}
\bibitem{Melville:2019wyy}
S.~Melville and J.~Noller,
%``Positivity in the Sky: Constraining dark energy and modified gravity from the UV,''
Phys. Rev. D \textbf{101} (2020) no.2, 021502
[erratum: Phys. Rev. D \textbf{102} (2020) no.4, 049902]
%doi:10.1103/PhysRevD.101.021502
[arXiv:1904.05874 [astro-ph.CO]].
%50 citations counted in INSPIRE as of 16 Dec 2022


\end{thebibliography}
\end{document}